\documentclass[aps,prl,twocolumn,footinbib,superscriptaddress,longbibliography]{revtex4-2}
\usepackage{graphicx}
\usepackage{color}
\usepackage{bm}
\usepackage{braket}
\usepackage{amsmath,amssymb}
\usepackage{calc}
\usepackage[shortcuts]{extdash}
\usepackage{bbold}
\usepackage[cal=boondoxo]{mathalfa}
\allowdisplaybreaks
\usepackage[bookmarksnumbered,bookmarksopen]{hyperref}
\hypersetup{
   colorlinks=true,
   citecolor=blue,
}

\usepackage{url}
\usepackage[normalem]{ulem}

\usepackage{array}
\usepackage{dcolumn}
\newcolumntype {L}{>{$}l<{$}}
\newcolumntype {C}{>{$}c<{$}}
\newcolumntype {R}{>{$}r<{$}}
\newcolumntype {s}[1]{@{\hspace*{#1}}}
\newcolumntype {S}[1]{@{\extracolsep{#1}}}
\newcolumntype {e}{@{\extracolsep {0pt}}}
\newcolumntype {E}{@{\extracolsep \fill}}
\newcolumntype {n}{@{}}
\newcolumntype {d}[1]{D{.}{.}{#1}}

\newcommand* {\vek}[1]{{\bm{\mathrm{#1}}}}
\newcommand* {\vekc}[1]{{\bm{\mathcal{#1}}}}
\newcommand* {\kk}{\vek{k}}
\newcommand* {\gv}{\vekc{g}}
\newcommand* {\rr}{\vek{r}}

\newcommand* {\gsis}{C_i}

\newcommand* {\gsistis}{C_{i \times \theta}}

\let\myRe\Re
\let\myIm\Im
\renewcommand{\Re}{\myRe\mathrm{e}\,}
\renewcommand{\Im}{\myIm\mathrm{m}\,}

\graphicspath{{./FIG/}}

\begin{document}

\title{Gauge-invariant absolute quantification of electric and
magnetic multipole densities\\ in crystals}

\author{R. Winkler}
\affiliation{Department of Physics, Northern Illinois University,
DeKalb, Illinois 60115, USA}
\affiliation{Materials Science Division, Argonne National Laboratory,
Lemont, Illinois 60439, USA}

\author{U. Z\"ulicke}
\affiliation{MacDiarmid Institute, School of Chemical and Physical Sciences,
Victoria University of Wellington, PO Box 600, Wellington 6140, New
Zealand}

\date{\today}

\begin{abstract}
Electric and magnetic multipole densities in crystalline solids,
including the familiar electric dipole density in ferroelectrics and
the magnetic dipole density in ferromagnets, are of central
importance for our understanding of ordered phases in matter.
However, determining the magnitude of these quantities has proven to
be conceptually and technically difficult.  Here we present a
universally applicable approach, based on projection operators, that
yields gauge-invariant absolute measures for all types of electric
and magnetic order in crystals.  We demonstrate the utility of the
general theory using concrete examples of electric and magnetic
multipole order in variants of lonsdaleite and diamond structures.
Besides the magnetic dipole density in ferromagnets, we also
consider, e.g., the magnetic octupole density in altermagnets.  The
robust method developed in this work lends itself to be incorporated
into the suite of computational materials-science tools.  The
multipole densities can be used as thermodynamic state variables
including Landau order parameters.
\end{abstract}

\maketitle

Electric and magnetic multipole densities represent thermodynamic
states of crystalline matter that are of central importance for
fundamental physics and technological applications~\cite{sin05}.
Well-known examples include the electric dipole density (electric
dipolarization) in ferroelectric and pyroelectric media and the
magnetic dipole density (magnetization) in ferromagnetic and
pyromagnetic media \cite{voi10, nye85, new05, pap02}.  Higher-rank
magnetic multipole densities are characteristic features of
antiferromagnets and altermagnets \cite{win23}.  Similarly,
higher-rank electric multipole densities exist in antiferroelectric
media \cite{win23, rab13}.  If we change the thermodynamic state,
e.g., in a phase transition from a paramagnetic state to a
ferromagnetic state or by reverting the electric dipolarization in a
ferroelectric medium, the amount by which the state variables change
must be independent of the path taken, i.e., independent of the
history of the system \cite{pap02, rei98}.  The multipole densities
are prime candidates for order parameters in Landau's theory of
phase transitions \cite{lan5e} applied to electrically and
magnetically ordered systems.

These basic concepts are well-known from thermodynamics.  However,
in the past they have posed significant challenges in explicit,
quantitative theories for electric and magnetic multipole densities
in crystalline media.
For crystals, a naive definition of electric dipolarization as the
dipole moment per unit cell is unsatisfactory, it depends on the
arbitrary choice of the unit cell \cite{mar74, kin93, res94, res10}.
Similar difficulties arise, for example, with orbital magnetic
dipoles \cite{hir97} and higher multipole moments associated with
clusters of atoms \cite{suz17}.  These difficulties are closely
related to the well-known fact that, in a multipole expansion for
finite systems, higher-rank multipoles beyond the lowest-rank
nonvanishing multipole are ill-defined, as they depend on the arbitrary
choice of the origin $\rr=0$ \cite{jac99}. Important progress
towards addressing this issue was made by the modern theories of
electric dipolarization and magnetization that provide
gauge-invariant expressions for the electric dipolarization and
magnetization based on properties of Bloch functions, i.e.,
independent of individual atoms or ions constituting a crystal
structure~\cite{kin93, res94, res07, res10, van18}.  However, the
modern theories still have a number of limitations.  Their
quantification of the electric dipolarization works only for
insulators~\cite{res02} and only pertains to polarization
\emph{changes} that must be evaluated along a continuous
all-insulating path connecting the initial and final state.
Furthermore, the modern theories define polarization only up to an
integer multiple of a polarization quantum.  When applied to media
with a spontaneous polarization, the choice of a reference state is
not trivial \cite{ber01, dre16}.  Most importantly, the extension of
the modern theories to higher-rank multipole densities beyond
dipolar order turns out to be difficult~\cite{ono19, dai20}.

Symmetry can be used to develop a robust quantitative
gauge-invariant theory of electric and magnetic multipole densities
in crystals, as we show in the following.  Key is a formulation not
in position space, where the origin $\rr=0$ is arbitrary, but in
reciprocal space (reciprocal lattice vectors $\gv$), where the
origin $\gv=0$ is uniquely defined. (Formally, $\gv=0$ corresponds
to the translationally invariant irreducible representations (IRs)
of the respective space group \cite{lud96}.)  Similar to the modern
theories \cite{res10, van18}, the formulation in reciprocal space
emphasizes that multipole densities represent macroscopic properties
of a crystal structure that cannot be associated with individual
atoms.  The following theory formalizes and quantifies the general
concepts underlying the case studies of multipole densities and
crystal order in variants of lonsdaleite and diamond that were
presented in Ref.~\cite{win23}.
Throughout, we use the term ``multipole densities'' to denote
macroscopic quantities in crystals \cite{win23}, as compared with
localized multipoles associated with individual atoms or clusters of
atoms in a crystal \cite{kur08, suz17}.  Note also that surface
effects do not contribute to these bulk properties.

Our formalism is based on a general systematic theory of
multipoles based on group theory, where spherical multipoles
represent quantities that transform irreducibly under the rotation
group $SO(3)$ \cite{ros55, win23}.  It is this generalized
definition of multipoles, which transcends the way electric and
magnetic multipoles are introduced by electrodynamics \cite{jac99},
that we use in the present work.  We illustrate below for the case
of electric dipole densities (rank $\ell =1$) that the group-theoretical
definition of multipoles yields results that are in line with the modern
theory of polarization.

In crystalline solids, lattice-periodic quantities $q(\rr)$ can be
equivalently expressed via their Fourier components $q(\gv)$ in
reciprocal space with reciprocal lattice vectors $\gv$.  Here
$q(\gv)$ may represent, e.g., scalar quantities such as the charge
density $\rho(\gv)$ or tensorial quantities such as a magnetization
density $\vek{m}(\gv)$.  If $G$ is the symmetry group of the
crystal, observable quantities $q(\gv)$ must transorm according to
the identity IR $\Gamma_1$ of $G$ \cite{misc:koster}.  If $G$ is not
the symmetry group of the crystal, as discussed in more detail
below, parts of $q(\gv)$ may transform according to different IRs
$\Gamma_\alpha$ of $G$.  To identify these parts, we can project
$q(\gv)$ onto the IRs $\Gamma_\alpha$ of the symmetry group
$G$~\cite{lud96}
\begin{subequations}
  \label{eq:project:f}
  \begin{equation}
    q_\alpha (\gv) = \Pi_\alpha\, q(\gv)
    \equiv \frac{n_\alpha}{h} \sum_{g \in G} \chi_\alpha(g)^\ast\,
    q [P(g)\gv] \; ,
  \end{equation}
  such that $q_\alpha (\gv)$ transforms according to
  $\Gamma_\alpha$.  Here $\chi_\alpha (g)$ are the characters of the
  IR $\Gamma_\alpha$ and $P(g)$ are the symmetry operators
  corresponding to the group elements $g \in G$, $n_\alpha$ is the
  dimension of $\Gamma_\alpha$, and $h$ is the order of $G$.
  This yields the decomposition
  \begin{equation}
    q(\gv) = \sum_\alpha q_\alpha (\gv) \; ,
  \end{equation}
\end{subequations}
because $\Pi_\alpha$ are orthogonal projection operators
\begin{subequations}
  \begin{equation}
    \label{eq:project:orthogonal}
    \Pi_\alpha \, \Pi_{\alpha'} = \delta_{\alpha\alpha'} \, \Pi_\alpha
  \end{equation}
  that obey the completeness relation~\cite{lud96}
  \begin{equation}
    \label{eq:project:complete}
    \sum_\alpha \Pi_\alpha = \openone \; .
  \end{equation}
\end{subequations}
For conceptual clarity, we use in Eqs.\ (\ref{eq:project:f}) the
coarse-grained projection operators $\Pi_\alpha$ that do not
distinguish between individual components of multi-dimensional IRs.
Group theory also defines fine-grained projection operators that
project onto individual components of multi-dimensional IRs
\cite{lud96}.

Often a projection (\ref{eq:project:f}) may show that a quantity
$q(\gv)$ transforms according to only one IR
$\Gamma_{\tilde{\alpha}}$ so that $q(\gv) = q_{\tilde{\alpha}}
(\gv)$.  If $G$ is the crystallographic point group of a crystal
structure, an observable quantity such as the charge density
$\rho(\gv)$ must transform according to the identity IR $\Gamma_1$
of $G$ \cite{misc:dipol}.  However, $G$ need not be the symmetry
group of the system.  For example, if $G$ is the parent point group
above the critical temperature $T_c$ of a ferroelastic or
ferroelectric phase transition and $q(\gv)$ represents the charge
density $\rho(\gv)$, the projection (\ref{eq:project:f}) becomes a
\emph{crystallographic multipole expansion} that allows one to
evaluate quantitatively the crystallographic multipole densities
$\rho_\alpha (\gv)$ arising in such ferroic phase transitions.
Above $T_c$, the entire charge density $\rho(\gv) = \rho_1 (\gv)$
transforms according to the trivial IR $\Gamma_1$ of the parent
point group $G$, whereas below $T_c$ (when the new symmetry group
$U$ is a subgroup of $G$ \cite{lan5e}), some fractions $\rho_\alpha
(\gv)$ of the total density transform according to nontrivial IRs
$\Gamma_\alpha$ of $G$ (say, an electric dipole density in
ferroelectric transitions, and an electric quadrupole density in
ferroelastic transitions).  The correspondence between the
crystallographic multipoles and the more familiar spherical
multipoles has been tabulated by Koster \emph{et al.}  \cite{kos63}.
It is exploited in the examples given below.

The quantities $q_\alpha (\gv)$ provide a complete characterization
of the respective multipole density they represent.  We can estimate
the magnitude of $q_\alpha (\gv)$ by evaluating, for example
\begin{equation}
  Q_\alpha = \sum_{\gv} \; |q_\alpha (\gv)| \; .
\end{equation}
Changing the origin $\rr=0$ in real space changes the phase of the
Fourier components $q_\alpha (\gv)$ in reciprocal space.  But it
does not affect their magnitude, so that a quantity like $Q_\alpha$
is, indeed, independent of the origin $\rr=0$ \cite{bri88}.

A high-symmetry parent crystal structure, though often illuminating,
is not needed to define the projection operators $\Pi_\alpha$; such
a structure may not always exist, not even hypothetically.
Generally, to identify magnetic or odd\=/$\ell$ electric multipole
densities in a system with point group $U$, we need to project onto
the IRs of the supergroup $G = U \times \gsistis$, where $\gsistis =\{e,
i, \theta, i\theta\}$ is the four-element group formed from space
inversion $i$ and time inversion $\theta$, with $e$ denoting the neutral
element.  To identify even\=/$\ell$ electric multipole densities
arising in a ferroelastic phase transition, we need to project onto
the IRs of the crystallographic point group that is realized when
the elastic deformation is zero.
A complete classification of the group-subgroup relations for
ferroic phase transitions is implicitly contained in the tabulation
of Aizu species \cite{aiz70, aiz79}.

\begin{table}
  \caption{\label{tab:dip-wurtzite} Electric dipole density $Q_2^-$ in hexagonal noncentrosymmetric wurtzite semiconductors
  ($C_{6v}$).  For comparison, the bottom row
  gives the Berry phase $\Delta\braket{\Phi}$ that quantifies the
  electric dipole density according to the modern theory of
  polarization~\cite{res10, van18}. Both quantities were calculated
  from the tight-binding model in Ref.~\cite{kob83}.}
  \renewcommand{\arraystretch}{1.2}
  \let\mc\multicolumn \tabcolsep 0pt
  \begin{tabular*}{1.0\linewidth}{nLE*{5}{d{3}}n}
    \hline \hline \rule{0pt}{2.8ex}
    & \mc{1}{c}{ZnO} & \mc{1}{c}{AlN} & \mc{1}{c}{CdS} & \mc{1}{c}{CdSe}
    & \mc{1}{c}{ZnS} \\ \hline
    Q_2^- & 1.66 & 1.36 & 2.25 & 2.14 & 2.06 \\
    \Delta \braket{\Phi} & 1.84\pi & 1.56\pi & 2.48\pi & 2.37\pi & 2.28\pi
    \\ \hline\hline
  \end{tabular*}

  \vspace{2ex}
  \caption{\label{tab:oct-zincblende} Electric octupole density
  $Q_2^-$ in cubic noncentrosymmetric zincblende semiconductors
  ($T_d$) calculated from the tight-binding model in
  Refs.~\cite{cha75, cha77}.}
  \begin{tabular*}{1.0\linewidth}{nLE*{7}{C}n}
    \hline \hline \rule{0pt}{2.8ex}
    & \mc{1}{c}{GaP} & \mc{1}{c}{GaAs}  & \mc{1}{c}{GaSb}
    & \mc{1}{c}{InP} & \mc{1}{c}{InAs}  & \mc{1}{c}{InSb}
    & \mc{1}{c}{ZnSe} \\ \hline
    Q_2^-
    & 0.471 & 0.443 & 0.396 & 0.608 & 0.585 & 0.466 & 0.940 \\ \hline\hline
  \end{tabular*}
\end{table}

\begin{figure}[t]
  \centering
  \includegraphics[width=0.99\linewidth]{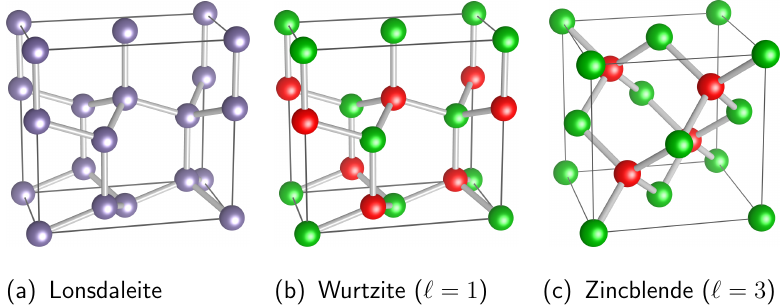}
  \caption[]{\label{fig:dip-wurtzite} (a) Crystal structure of
  lonsdaleite (point group $D_{6h}$). (b) Crystal structure of
  wurtzite ($C_{6v}$), which hosts an electric-dipole density
  (multipole rank $\ell=1$).  (c) Crystal structure of zincblende
  ($T_d$), which has an electric-octupole density ($\ell=3$).}
\end{figure}

As illustrative examples, Tables~\ref{tab:dip-wurtzite}
and~\ref{tab:oct-zincblende} list the electric dipole density in
wurtzite semiconductors and the electric octupole density in
zincblende semiconductors, respectively, calculated using the
$s$-$p$ tight-binding (TB) models in Refs.~\cite{kob83}
and~\cite{cha75, cha77}.  See Fig.~\ref{fig:dip-wurtzite}(b) for an
illustration of the hexagonal noncentrosymmetric wurtzite structure
that is realized by common III\=/V and II\=/VI semiconductors
including AlN, ZnO, and CdSe.  Ignoring time inversion symmetry, the
point group of wurtzite is $C_{6v}$ that permits an electric dipole
density \cite{voi10, kos63, nye85}.  Following the formalism
described above, we introduce $C_{6v} \times \gsis = D_{6h}$ as the
parent supergroup of wurtzite, where $\gsis = \{e,i\}$.  An electric
dipole density transforms according to the IR $\Gamma_2^-$ of
$D_{6h}$ (Koster's notation \cite{kos63}), i.e., a dipole density is
forbidden under $D_{6h}$.  But when the symmetry is reduced from
$D_{6h}$ to $C_{6v}$, the IR $\Gamma_2^-$ of $D_{6h}$ is mapped onto
the IR $\Gamma_1$ of $C_{6v}$ \cite{kos63} so that an electric
dipole density becomes symmetry-allowed in wurtzite.  Using the TB
model in Ref.~\cite{kob83}, we can evaluate the charge density
$\rho(\gv)$ in several wurtzite semiconductors and project
$\rho(\gv)$ onto the IRs $\Gamma_1^+$ and $\Gamma_2^-$ of $D_{6h}$,
yielding partial charge densities $\rho_1^+ (\gv)$ and $\rho_2^-
(\gv)$ \cite{misc:tb}.  From the latter, we obtain the dipole
densities $Q_2^-$ listed in Table~\ref{tab:dip-wurtzite}.  The
quantity $Q_2^-$ is a gauge-invariant measure for the electric
dipole density in these materials.  As to be expected, the
projection of $\rho(\gv)$ on all other nontrivial IRs of $D_{6h}$
vanishes, i.e., $\rho(\gv) = \rho_1^+ (\gv) + \rho_2^- (\gv)$.

\begin{figure}[t]
  \centering
  \includegraphics[width=0.51\linewidth]{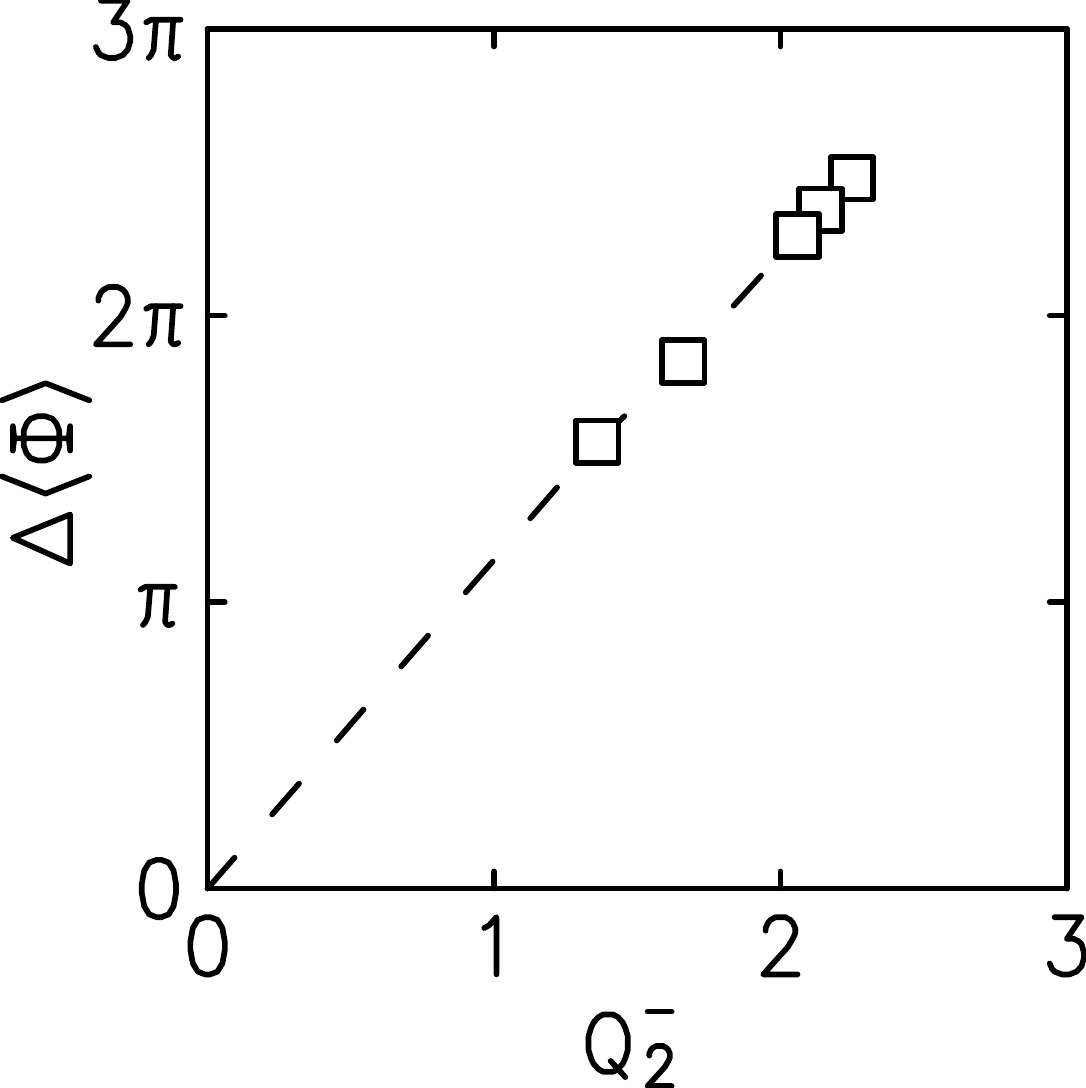}
  \caption[]{\label{fig:dip-berry} Berry phase $\Delta
  \braket{\Phi}$ versus the dipole density $Q_2^-$ for the wurtzite
  semiconductors in Table~\ref{tab:dip-wurtzite}.  The dashed line
  is a guide to the eye.}
\end{figure}

For comparison, Table~\ref{tab:dip-wurtzite} also gives the Berry
phase
\begin{subequations}
  \label{eq:berry}
  \begin{equation}
    \Delta \braket{\Phi}
    = \int_0^1 d\lambda \, \frac{d \braket{\Phi (\lambda)}}{d\lambda}
  \end{equation}
  with
  \begin{equation}
   \label{eq:berryDetail}
    \braket {\Phi} =
    \Im \sum_n \int \frac{d^3k}{\Omega} \,
    \braket{u_{n\kk} | \partial k_z | u_{n\kk}}
  \end{equation}
\end{subequations}
that represents the electric dipole density according to the modern
theory of polarization \cite{res10, van18, misc:nuclear}.  Here,
$\lambda$ parameterizes a path in parameter space that connects
dipolar wurtzite with nonpolar lonsdaleite as a reference structure
\cite{misc:lambda}. See Fig.~\ref{fig:dip-wurtzite}(a) for an
illustration of the centrosymmetric lonsdaleite crystal structure,
which has the crystallographic point group $D_{6h} = C_{6v} \times
\gsis$, i.e., lonsdaleite does not permit odd\=/$\ell$ electric
multipole densities.  In Eq.~(\ref{eq:berryDetail}), $u_{n\kk}$
denotes the lattice-periodic part of the Bloch function for band
$n$, and the sum runs over all occupied bands.  The $z$ axis points
along the principal axis of wurtzite, and $\Omega$ is the
cross-sectional area of the Brillouin zone perpendicular to the $z$
axis.
The dipole density $Q_2^-$ and the Berry phases $\Delta
\braket{\Phi}$ for the different materials considered in
Table~\ref{tab:dip-wurtzite} are evidently proportional to each
other [Fig.~\ref{fig:dip-berry}].  Thus, even though the microscopic
theories underlying $Q_2^-$ and $\Delta \braket{\Phi}$ are very
different, these quantities represent the same physics.

The cubic noncentrosymmetric zincblende structure, realized by
common III\=/V and II\=/VI semiconductors including GaAs and ZnSe,
has the point group $T_d$ that permits an electric octupole density
\cite{kos63, win23}.  The parent supergroup of zincblende is $T_d
\times \gsis = O_h$, and an electric octupole density transforms
according to the IR $\Gamma_2^-$ of $O_h$ (which is mapped onto the
IR $\Gamma_1$ of $T_d$, as to be expected \cite{kos63}).  Using the
TB model in Refs.~\cite{cha75, cha77}, we calculate partial charge
densities $\rho_1^+ (\gv)$ and $\rho_2^- (\gv)$ that result in the
octupole densities $Q_2^-$ for particular zincblende materials
listed in Table~\ref{tab:oct-zincblende} \cite{misc:soc}.  These
quantities are a gauge-invariant measure for the electric octupole
density in these materials.  Again, the projection of $\rho(\gv)$ on
all other nontrivial IRs of $O_h$ vanishes, i.e., $\rho(\gv) =
\rho_1^+ (\gv) + \rho_2^- (\gv)$.

\begin{figure}[t]
  \centering
  \includegraphics[width=0.70\linewidth]{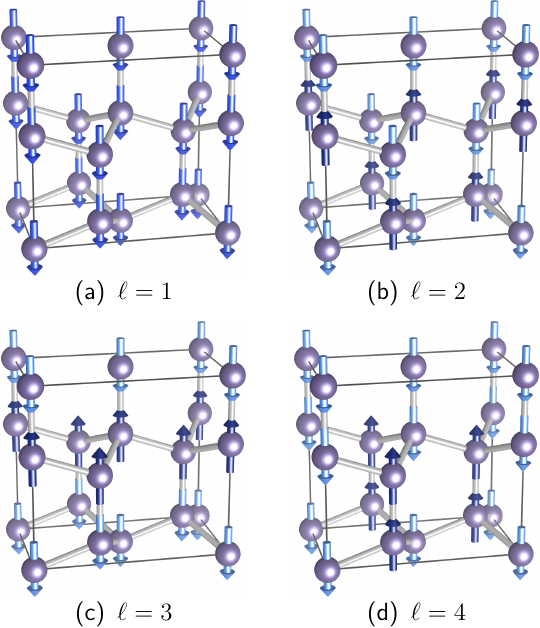}
  \caption[]{\label{fig:lons-magnet} Magnetic multipole densities in
  the lonsdaleite family.  Local magnetic moments give rise to (a) a
  magnetization ($\ell = 1$), (b) a quadrupolarization ($\ell = 2$),
  an octupolarization ($\ell = 3$), and a hexadecapolarization
  ($\ell = 4$).}
\end{figure}

\begin{table}
  \caption{\label{tab:mag-wurtzite} Magnetic multipole densities
  $S_z$ of rank $\ell$ in lonsdaleite semiconductors ($D_{6h}$)
  calculated from the tight-binding model in Ref.~\cite{kob83}; see
  Fig.~\ref{fig:lons-magnet}.  The second column gives the IR of the
  multipole densities under $D_{6h}$.  The third column gives the
  point group $U$ (subgroup of $D_{6h}$) realized with the
  respective multipole density.}
  \let\mc\multicolumn \tabcolsep 0pt
  \begin{tabular*}{1.0\linewidth}{nLECCd{4}d{5}d{5}n}
    \hline \hline \rule{0pt}{2.8ex}
    & D_{6h} & U & \mc{1}{c}{C} & \mc{1}{c}{Si} & \mc{1}{c}{Ge} \\ \hline \rule{0pt}{2.7ex}
    \ell = 1 & \Gamma_2^+ & C_{6h} & \mc{1}{C}{< 10^{-8}} & \mc{1}{C}{2.1 \times 10^{-6}} & \mc{1}{C}{1.4 \times 10^{-4}} \\
    \ell = 2 & \Gamma_1^- & D_6    & 0.143  & 0.299 & 0.289 \\
    \ell = 3 & \Gamma_3^+ & D_{3d} & 0.0362 & 0.0827 & 0.0939 \\
    \ell = 4 & \Gamma_4^- & D_{3h} & 0.102  & 0.230 & 0.237 \\ \hline\hline
  \end{tabular*}
\end{table}

As an exemplary model calculation, we also evaluate the magnetic
multipole densities with ranks $\ell=1$ to $4$ for the magnetic
variants of lonsdaleite shown in Fig.~\ref{fig:lons-magnet},
starting from the TB Hamiltonian in Ref.~\cite{kob83} and using the
TB parameters from Ref.~\cite{cha77} complemented by local exchange
energies of $0.1$~eV arranged as in Fig.~\ref{fig:lons-magnet}.
Here the magnetic multipole densities are derived from the spin
magnetization density $\vek{s} (\gv)$.  For the collinear magnetic
systems in Fig.~\ref{fig:lons-magnet}, the spin density becomes $s_z
(\gv) = \rho_\uparrow (\gv) - \rho_\downarrow (\gv)$, where
$\rho_\sigma (\gv)$ is the charge density due to electrons with spin
orientation $\sigma = {\uparrow}, \downarrow$ [with $\rho(\gv) =
\rho_\uparrow (\gv) + \rho_\downarrow (\gv)$].  To relate with
Koster's tabulation of IRs \cite{kos63}, we project onto the IRs of
$D_{6h}$ instead of the IRs of $D_{6h} \times \{e,\theta\}$.  In
each case, this reveals that the entire spin density $s_z (\gv)$
transforms according to only one IR $\Gamma_\alpha$ listed in
Table~\ref{tab:mag-wurtzite}.  For the different materials, we then
list in Table~\ref{tab:mag-wurtzite} the spin density
$S_z = \sum_\gv |s_z (\gv)|$ \cite{misc:mag-atom}.

Commonly, collinear antiferromagnetic systems as in
Figs.~\ref{fig:lons-magnet}(b)-(d) are characterized by a N\'eel
vector.  Clearly, the different magnetic multipole densities $\ell >
1$ in Table~\ref{tab:mag-wurtzite} provide a more fine-grained
characterization of these systems.  As discussed in
Ref.~\cite{win23}, the magnetic octupole density (rank $\ell=3$) in
Fig.~\ref{fig:lons-magnet}(c) is typical for altermagnets that have
recently been attracting interest \cite{sme22, sme22a}.  The
variants of lonsdaleite permitting even\=/$\ell$ magnetic multipole
densities are magnetoelectric.

In conclusion, our theory has several advantageous properties.
It provides a unified, physically transparent theory of electric and
magnetic multipole densities in crystalline media, treating all
these quantities on the same footing.  It is in line with
established phenomenological (group-theory based) studies of
electric dipolarization and magnetization in crystalline media
\cite{misc:dipol} and also with the modern theory of
polarization~\cite{kin93, res94, res07, res10, van18}.  Beyond the
dipolar order in ferroelectric and ferromagnetic media, our approach
can also quantify the unconventional, higher-rank electric and
magnetic order present in the materials currently attracting
greatest interest, including altermagnets, antiferromagnets and
magnetoelectrics.
The projection (\ref{eq:project:f}) is valid for insulators,
semimetals, and metals.
The projected multipole densities $q_\alpha (\gv)$ and $Q_\alpha$
are gauge-invariant quantities with absolute values independent of a
reference state.  There is no ambiguity arising from a polarization
quantum in our theory.  Therefore, these quantities are well-suited
as intensive state variables for thermodynamic theories \cite{rei98}
including the order parameters in Landau's theory of phase
transitions \cite{lan5e}.
If distortions gives rise to multiple multipole densities $q_\alpha
(\gv)$ as in multiferroics, these can be evaluated independently.
Magnetic multipole densities can be evaluated for collinear and noncollinear
magnetic order.
Our robust theory lends itself to be incorporated into the suite of
computational materials-science tools, thus complementing current
approaches that focus on the modern theories \cite{res10} and local
multipole configurations~\cite{spa13, suz18, pic24}.

\begin{acknowledgments}
  Work at Argonne was supported by DOE BES under Contract No.\
  DE-AC02-06CH11357.  We thank R.~Resta for correspondence
  that prompted us to correct the description of our formalism.
  RW benefited from discussions with J.~Rau.
\end{acknowledgments}


\begin{thebibliography}{46}
\bibitem [{misc:koster()}]{misc:koster}%
  \BibitemOpen
  \bibinfo {note} {Throughout this work, we follow Koster's well-established
  designation of irreducible representations (IRs) for the crystallographic
  point groups \cite{kos63}. The identity IR is denoted $\Gamma_1$, unless the
  group contains space inversion as a symmetry element, when this IR is
  denoted~$\Gamma_1^+$.}\BibitemShut {Stop}%
\bibitem [{misc:dipol()}]{misc:dipol}%
  \BibitemOpen
  \bibinfo {note} {Early studies used the concept of invariance to identify the
  range of point groups of dipolar media. A material with crystallographic
  point group $G$ can be ferroelectric or pyroelectric if under $G$ an electric
  dipolarization transforms according to the identity IR $\Gamma_1$ of $G$
  \cite{voi10, nye85}. Similarly, a material can be ferromagnetic or
  pyromagnetic if under $G$ a magnetization transforms according to $\Gamma_1$
  of $G$ \cite{tav58}. More recently, the range of crystal classes permitting
  electric dipolarizations and magnetizations have been tabulated, e.g., in
  Ref. \cite{new05}.}\BibitemShut {Stop}%
\bibitem [{misc:tb()}]{misc:tb}%
  \BibitemOpen
  \bibinfo {note} {The Fourier representation $q(\gv)$ is simplified in TB
  models, where the real-space functions $q(\rr)$ are sums of $\delta$
  functions.}\BibitemShut {Stop}%
\bibitem [{misc:nuclear()}]{misc:nuclear}%
  \BibitemOpen
  \bibinfo {note} {The nuclear contribution to electric multipole densities,
  typically included in the modern theory of electric dipolarization
  \cite{res10}, does not arise within the TB model studied here. However, our
  formalism could also account for the nuclear contributions to the charge
  density $\rho(\gv)$ in a suitably extended model.}\BibitemShut {Stop}%
\bibitem [{misc:lambda()}]{misc:lambda}%
  \BibitemOpen
  \bibinfo {note} {For the materials in Table~\ref{tab:dip-wurtzite}, the path
  $\lambda$ in Eq.\ (\ref{eq:berry}) connects dipolar wurtzite with nonpolar
  lonsdaleite as a reference structure. The TB parameters for the latter
  structures are weighted averages of the TB parameters for the anion and
  cation of the respective wurtzite structures, chosen such that the system
  remains insulating along the path $\lambda$. This is conceptually analogous
  to the layered hexagonal structure considered as reference structure for
  wurtzite in Ref.~\cite{dre16}. Note that both lonsdaleite and the layered
  hexagonal structure have the space group No.~194, $P6_3/mmc$.}\BibitemShut
  {Stop}%
\bibitem [{misc:soc()}]{misc:soc}%
  \BibitemOpen
  \bibinfo {note} {The TB model in Ref.~\cite{cha77} takes into account
  spin-orbit coupling (SOC) whereas the model in Ref.~\cite{kob83} ignores SOC.
  However, SOC has only a marginal effect on the octupole densities in
  Table~\ref{tab:oct-zincblende}.}\BibitemShut {Stop}%
\bibitem [{misc:mag-atom()}]{misc:mag-atom}%
  \BibitemOpen
  \bibinfo {note} {Commonly a macroscopic magnetic dipole density ($\ell = 1$)
  is interpreted as ``magnetic moment per (magnetic) atom.'' However, this is
  not meaningful for magnetic multipole densities with rank $\ell > 1$ that
  cannot be attributed to individual atoms.}\BibitemShut {Stop}%
\end{thebibliography}
\end{document}